# Near-field imaging of surface-plasmon vortex-modes around a single elliptical nanohole in a gold film

Claudia Triolo[1*], Salvatore Savasta [1,2], Alessio Settineri[1], Sebastiano Trusso[3*], Rosalba Saija[1], Nisha Rani Agarwal[4], Salvatore Patanè[1]

[1] *Dipartimento di Scienze Matematiche e Informatiche, Scienze Fisiche e Scienze della Terra, University of Messina, Messina, Italy*
[2] *Theoretical Quantum Physics Laboratory, Cluster for Pioneering Research, RIKEN, Wako-Shi, Saitama, 351-0198, Japan*
[3] *CNR-IPCF, Istituto per i Processi Chimico-Fisici del CNR, Messina, Italy*
[4] *Faculty of Science, University of Ontario Institute of Technology, Oshawa, ON Canada*

*Corresponding authors: trusso@me.cnr.it, trioloc@unime.it

## Abstract

We present scanning near-field images of surface plasmon modes around a single elliptical nanohole in 88 nm thick Au film. We find that rotating surface plasmon vortex modes carrying extrinsic orbital angular momentum can be induced under linearly polarized illumination. The vortex modes are obtained only when the incident polarization direction differs from one of the ellipse axes. Such a direct observation of the vortex modes is possible thanks to the ability of the SNOM technique to obtain information on both the amplitude and the phase of the near-field. The presence of the vortex mode is determined by the rotational symmetry breaking of the system. Finite element method calculations show that such a vorticity originates from the presence of nodal points where the phase of the field is undefined, leading to a circulation of the energy flow. The configuration producing vortex modes corresponds to a nonzero total topological charge (+1).

Subwavelength holes play an important role in several advanced techniques in nano-optics. Their optical behavior is central to high resolution near-field scanning microscopies[1-7], extraordinary optical transmission phenomena[8], surface-plasmon assisted light beaming[9,10], fluorescence correlation spectroscopy[11], and optical trapping[12]. The theory of light diffraction through an aperture is very complex even in the simplest geometries and, usually, it is possible to distinguish two general cases[13]: an aperture with a radius $r$ significantly larger than the wavelength of the impinging radiation ($r \gg \lambda_0$), that can be treated using the Huygens-Fresnel principle; and an aperture with a radius $r$ significantly smaller than the wavelength of the impinging radiation ($r \ll \lambda_0$), discussed for the first time by Bethe in 1944[14], that provided a complete solution of the Maxwell's equations. When the lateral dimensions of the aperture is smaller than half the wavelength, light cannot propagate through the hole and the transmission is typically very weak. However, a real subwavelength aperture[15] behaves very differently because the finite thickness and the finite conductivity of the metal can give rise to exotic optical phenomena, related to the electromagnetic field distribution around the aperture, to the excitation of surface plasmon polaritons (SPPs), and manifestations of spin-orbit interactions (SOI)[16,17]. Recently, there has been a great interest in understanding and demonstrating SOI of light[18]. At the subwavelength scale, SOI phenomena can be very relevant since they allow to control the spatial degrees of freedom of light (intensity distribution of electromagnetic field and propagation paths) selecting the spin states of incident photons ($\sigma = \pm 1$ that corresponds to the right- and left-hand circular polarization of light). SOI results from the interaction between light carrying spin (as circularly polarized light) and spatially inhomogeneous or anisotropic materials. This interaction leads to many interesting phenomena, such as the photonic spin Hall effect (SHE)[19-21], optical spin-to-orbital angular momentum conversion[22,23], generation of spin-dependent light force[24], plasmonic vortex modes[25,26], and spin-momentum locking effect in evanescent waves[27,28]. However, due to the small momentum carried by photons, the SOI of light are exceedingly small and their experimental observation is challenging [29-33]. In order to explore such weak processes, plasmonic metamaterials (nanoparticles or nanostructured thin films that support plasmon resonances) are largely used, thanks to the flexibility of their structural design and the enhanced subwavelength local field. Strong spin Hall effects have been observed in metasurfaces with anisotropic nanostructures[26,34]. Considering circularly-polarized incident light, helicity-dependent transverse motion of refracted light has been observed in such systems[35,36]. Moreover, for linearly polarized incident light, it has been shown that the refracted beam can display a transverse splitting of the two spin components[16].

In this context, we study the SOI effects of an evanescent field scattered by an isolated elliptical nanohole in a 88 nm thick Au film by using a near-field scanning optical microscope (SNOM) working in transmission mode. Exploiting the rotational symmetry breaking due to the elongated shape of the nanohole, a plasmonic vortex mode is generated by illuminating the hole with an incident light beam without a spin state ($\sigma = 0$ corresponds to linearly polarized light). We present a direct observation of the vortex mode by SNOM measurements. This observation relies on the unique ability of this technique to provide information on both amplitude and phase of the electromagnetic near-field distribution[37]. Specifically, we demonstrate that the geometric anisotropy plays a key role in the near-field SOI and that it is responsible for the generation of a plasmonic vortex mode from a single elongated nanohole. This is a much simpler structure than those generally used to investigate these effects, such as plasmonic rings, Archimedes spirals and similar composite structures[38,39]. In addition, the rotation direction of the vortex (right- or left-hand rotation) depends on the angle $\theta$ between the polarization direction of the incident field and the symmetry axes of the ellipse, which induce a spin-dependent splitting in the scattered field and controls its spatial distribution in near-field.

**Results and Discussion**

In order to explore the contribution related to the scattered optical field around the elliptical nanohole to the SOI phenomena, the SNOM measurements are performed using two linear polarizers with crossed optical axes: the first (#1) placed in the optical path of the incident field before the sample (see Fig. 1) allows to polarize the incident beam; the second one (#2) filters the electromagnetic field collected by the probe after the interaction with the sample.

Figure 2a shows the SNOM image of the scattered field (scan area 5μm×5μm) performed using a linearly polarized incident beam at $\lambda_{exc} = 632$ nm with normal incidence. The polarization axis forms an angle of $\theta = +45°$ respect to the *x*-axis of the elliptical nanohole (its direction is indicated by the white dotted line). In order to suppress the contribution of the incident field, the second linear polarizer (#2, placed before the CCD camera) is crossed (rotated by 90°) respect to the first one. As will be confirmed below, with the help of numerical calculations, under these illumination and detection conditions, the electromagnetic field shows a spiral-like distribution with a counterclockwise rotation, corresponding to a surface plasmon vortex mode. In order to provide a complete explanation of this exotic distribution of the detected electromagnetic field, it is useful to understand the properties of the detected signal. The SNOM probe collects almost only the *z*-component of the total field generated close to the sample surface and converts it to a propagating wave (see Methods). The electromagnetic field intensity detected by the CCD camera can approximately be written as[37]:

$$I(\mathbf{r}) = |E_{bg} + \alpha E_z(\mathbf{r})|^2 \cong |E_{bg}|^2 + 2Re\left(\alpha E_{bg}^* \cdot E_z(\mathbf{r})\right), \qquad (1)$$

where $E_z(\mathbf{r})$ is the z-component of the electric field in the near-field region on the top of the sample and $E_{bg}$ is the position-independent background field that includes all the electromagnetic field scattered by the sample (also due disorder and interface roughness) or, more generally, from the sample-probe system in the direction orthogonal to the incident polarization. The incident electromagnetic field contributes to the background field, because the Au layer thickness is sufficiently thin to allow to a small fraction of the incident light to cross it. In Eq. (1), $\alpha$ is the transfer function of the SNOM probe, and we assume that the probe mainly collects the z-component of the electric field. The approximate expression in Eq. (1) is obtained assuming $|E_{bg}| \gg |E_z|$, where $|E_{bg}|^2$ is the background intensity, while $2Re\left(\alpha E_{bg}^* \cdot E_z(\mathbf{r})\right)$ is an interference term that depends on the probe position with respect to the sample and provides information on both the amplitude and the phase of the near-field component $E_z(\mathbf{r})$. This interference term can be both positive and negative with respect to the constant background component and it is responsible of the strong contrast observed close to the nanohole (see Fig. 2a).

In order to understand the origin of the spiral-like distribution of the evanescent electromagnetic field around the nanohole, FEM (Finite Element Method) simulations have been performed, considering the same experimental illumination conditions (orientation of the incident polarization and normal incidence of the exciting beam). Since the interesting optical behavior is ascribed to the near-field amplitude collected by the SNOM probe, and since we can assume both phase and amplitude of the background field as position independent, in our FEM simulations we set both these quantities to zero and evaluate a map of the z-component of the total electric field $Re(E_z)$. Notice that, a non-zero value of this phase can only affect the position of maxima and minima of the interference pattern, without changing the overall spiral shape and its rotation direction.

Figure 2b shows the numerical calculation of $Re(E_z)$ in a plane at a distance $h = 2$ nm from the sample surface. The resulting field distribution is in very good qualitative agreement with the experimental result shown in Fig. 2a. Figure 2c displays the distribution of $|E_z|^2$, which is characterized by an interference pattern with the maxima aligned along the incident polarization. However, the spiral structure shown in Figs. 2a and 2b is not visible. This different behavior confirms that the detected signal is described by Eq. (1). Figure 2d shows the map of the numerically calculated square modulus of the total electric field $|E|^2$, displaying an interference pattern very similar to that shown in Fig. 2c. In fact, the nanohole acts as a point like source of SPPs[40], which are more pronounced in the direction of electric field polarization since, along this direction, the incident electromagnetic field separates positive and negative charges in the metal and reinforces the external field around the aperture. This charge distribution close to the nanohole produces both LPRs and SPPs, propagating on the metal surface.

SNOM measurements and FEM simulations were repeated considering an incident light beam with polarization aligned with the major axis of the nanohole ($\theta = +90°$). The corresponding SNOM image is reported in Fig. 3a. It shows an interference pattern along the polarization direction of the incident light beam. A similar interference pattern is obtained by FEM simulations (see Figs 3b-d). The images in Figs. 3a and 3b are in good qualitative agreement only in the lower half of the nanohole plane. On the contrary, the theoretically predicted interference pattern is not visible in the upper half of the nanohole plane of the SNOM image. This absence of interference on one half-plane could be explained as the disturbance induced by a nearby gold nanoparticle (a side effect of the nano-drilling process) clearly visible in the AFM image (see Methods). We observe that, with this incident linear polarization, no spiral-like field distribution is visible in both SNOM and calculated maps. The results shown in Figs. 2 and 3 agree with previous theoretical analysis and calculations[31], where it has been shown that SP vortex modes, carrying extrinsic orbital angular momentum, can be induced under linearly polarized illumination of individual anisotropic nanostructures, when the polarization direction differs from the symmetry axes of the nanostructure. This effect is a manifestation of SOI of light and it is intimately linked to the prediction and observation of a related effect: the splitting of light trajectories depending on the spin state of the incident light[18].

A deeper understanding of the origin of the vortex-modes under illumination with linearly-polarized light can be obtained by calculating the phase distribution of the longitudinal component $E_z$ of the electric field near the surface. Figure 4 shows the spatial distribution of the phase of the z-component of the electric field and of the optical momentum $\boldsymbol{p}_0$[27] around the nanohole, calculated by FEM simulations both for a $\theta = +45°$ (a,c) and $\theta = +90°$ (b,d) incident polarization direction. The optical momentum $\boldsymbol{p}_0$ can be written in terms of the electric $\boldsymbol{E}$ and magnetic $\boldsymbol{H}$ components of the optical field as[26,27]:

$$\boldsymbol{p}_0 = \frac{\gamma}{2} Im[\boldsymbol{E}^* \cdot (\nabla)\boldsymbol{E} + \boldsymbol{H}^* \cdot (\nabla)\boldsymbol{H}] \qquad (2)$$

where $\gamma = (8\pi\omega)^{-1}$ is a constant in Gaussian units that depends on the frequency $\omega$ of the incident light. Both panels 4a and 4b show the presence of nodal points (or phase singularities), where the phase is undefined. These points correspond to places where the intensity of the field $|E_z|^2$ is zero and in general all the phase values from 0 to $2\pi$ occur around these singularities, leading to a circulation of the optical energy[35,41].

These points, which are extremely general features of optical fields, can be observed in the plane orthogonal to the propagation direction of beams carrying orbital angular momentum $\boldsymbol{p}_0$[42]. In general, they can occur in a plane when three or more waves interfere[41]. Here the elliptical hole can be viewed as the combination of two dipole-like oscillators of different lengths, and the phase pattern is the result of the interference between these two dipoles and the SPPs propagating on the gold surface. These phase singularities do not occur for circular holes (see Supplementary Information). We can express the $E_z$ component of the electric field as $E_z(\mathbf{r}) = \rho(\mathbf{r}) \exp[i\,\chi(\mathbf{r})]$. By defining a circuit $C$ enclosing the point where $E_z$ vanishes (both the real and the imaginary part), the strength of the singularity, or topological charge, can be calculated as[41]:

$$\ell = \frac{1}{2\pi} \oint d\chi, \tag{3}$$

where $d\chi$ is the phase variation along the circuit $C$ and $\ell$ is the topological charge, which is positive if the phase increases in the clockwise direction. The contribution of $E_z$ to the electromagnetic energy flow can be expressed as $\mathbf{j} = E_z^* \nabla E_z = \rho(\mathbf{r}) \nabla \chi$. The resulting energy flux is thus oriented in the direction of the phase change $\nabla \chi$, and the phase singularities are therefore vortices of the optical energy flow, named optical vortices[41]. Hence, these phase-singularity points determine the overall flow of the optical energy in a plane. As observed in several studies[41-43], these, and other phase singularities (where $\nabla \chi = 0$) organize the global spatial structure of the optical field. They constitute a *skeleton* on which the phase and intensity structure hangs. For example, a phase singularity with a helical phase around it, generating a circulation of the local momentum density (phase gradient), produces an orbital angular momentum.

When the incident polarization is collinear with one of the hole axes (Fig. 4a), no nodal points are visible inside and on the hole edges. The closest vortex singularities are four with topological charges $\ell = \pm 1$ and are located at about 260 nm from the center. It is interesting to notice that they compensate, so that, the total topological charge $\ell_{tot} = \sum_{i=1}^{4} \ell_i = 0$.

Specifically, the top left point has a charge $\ell_1 = -1$, the top right charge is $\ell_2 = +1$, the bottom right: $\ell_3 = -1$, and the bottom left has a charge $\ell_4 = +1$. Figure 4c, including the associated energy flow $\mathbf{j}$, clearly shows the relationship between the charge sign and the rotation sense of the current around each point. The absence of an overall surface-plasmon vortex-mode under this incident polarization condition is closely related to the observed compensation of the topological charges. Notice that also singular points at larger distances display this kind of charge compensation, otherwise absent in the phase maps obtained with linearly polarized incident light (Figs. 4b, d). When the linear polarization direction of the incident light is rotated with respect to the major axis of the elliptical nanohole ($\theta = +45°$), a more complex phase map, with a larger number of optical vortices appears (see Figure 4b). Specifically, the phase singularities are placed on a band along the direction orthogonal to the incident polarization. One of these points is located at the center of the nanohole with $\ell_0 = -1$ and two are located on the edge of the hole, with topological charges $\ell_1 = \ell_2 = +1$. Hence, the surface surrounded by an imaginary line around the nanohole, shows an unbalanced total charge $\ell_{tot} = \sum_{i=0}^{2} \ell_i = +1$. The SPP vortex mode experimentally observed (Fig. 2a) is intimately related to this overall charge unbalance that determines a net non-zero vorticity (see also Fig. 4d). Moreover, all the singular points beyond the edge constitutes pairs with opposite topological charges. Hence the total charge for this configuration is determined by the unbalanced total charge inside the surface surrounded by an imaginary line around the nanohole: $\ell_{tot} = \sum_{i=0}^{2} \ell_i = +1$. This arrangement of singular points, resulting from the hole-anisotropy and from the rotated polarization direction, determines a helix arrangement of arches of almost constant phase (see Fig. 4b), interrupted by abrupt phase changes (in correspondence of the singular points). This phase structure is closely linked to the helix shape which can be observed in the SNOM image in Fig. 2a, and to the FEM calculated $Re(E_Z)$ in Fig. 2b. As shown in Fig. 4d, this charge unbalance induces an overall vorticity of the energy flux $\boldsymbol{j}$, and hence a SPP mode with nonzero orbital angular momentum.

Figure 5 shows the spatial distribution (2μm × 2μm) of the angle $\beta$ that the spin density $\boldsymbol{s}$ creates with the $xy$-plane, calculated by FEM simulations at $h = 2$ nm from the Au surface. The spin density $\boldsymbol{s}$ can be written in terms of the electric $\boldsymbol{E}$ and magnetic $\boldsymbol{H}$ components of the optical field as[26,27]:

$$\boldsymbol{s} = \frac{\gamma}{2} Im[\boldsymbol{E}^* \times \boldsymbol{E} + \boldsymbol{H}^* \times \boldsymbol{H}] \qquad (4)$$

where $\gamma = (8\pi\omega)^{-1}$ is a constant in Gaussian units that depends on the frequency $\omega$ of the incident light. We considered two incident polarization directions: (a) coincident with the major axis of the elliptical nanohole ($\theta = +90°$), and (b) rotated by $\theta = +45°$. The two panels in Fig. 5 show as the spin density distribution is affected by the linear polarization. The distribution of the angle $\beta$ shows that the spin $\boldsymbol{s}$ is orthogonal to the $xy$-plane in correspondence to the phase singularities (see Fig. 4) and $\beta$ can be positive or negative (upwards or downwards with respect to the $xy$-plane) depending on the topological charge of each nodal point and, hence, on the rotation direction of the vorticity of the energy flux $\boldsymbol{j}$. Since in these phase-singularity points the optical momentum $\boldsymbol{p}_0$ possesses only components on the $xy$-plane, here the spin of the optical field turns out to be almost completely transverse to the optical momentum $\boldsymbol{p}_0$. This is a clear manifestation of the "spin-momentum locking" effect[26,42], due to the evanescent behavior of light in the near-field region.

**Conclusions**

We have reported scanning near-field images of surface plasmon modes around a single elliptical nanohole, which provide a direct evidence of rotating surface plasmon vortex modes. Specifically, we have demonstrated that surface plasmon resonances with orbital angular momentum can be induced under linearly polarized illumination. These vortex modes arise only when the incident polarization direction differs from one of the ellipse axes. Such a direct observation of the vortex mode was possible thanks to the ability of the SNOM technique to provide information on both the amplitude and the phase of the near-field, thus, giving an accurate description of the spatial structure of surface plasmon polaritons. By developing FEM calculations, we were able to reproduce the observed plasmonic vortex modes and to infer their origin. We have shown that such a vorticity originates from the presence of nodal points with an overall charge different from zero, leading to an overall circulation of the energy flow. Analyzing the spatial distribution of the spin density vector, we found that $\boldsymbol{p}_0$ and $\boldsymbol{s}$ are orthogonal to each other in correspondence of the phase singularities. This is a clear manifestation of the "spin-momentum locking" effect, due to the evanescent behavior of the light. The simple detection method here used can be applied to probe localized surface plasmons in a variety of individual or coupled nanoholes and nanostructures[44,45]. Furthermore, it would be interesting to study these effects on dielectric surfaces supporting surface polaritons[46,47]. Future research is needed to investigate polarization-dependent optical momenta and forces around these elliptical nanoholes and around the resulting nodal points[48]. These measurements could be performed by means of nano-cantilevers[27].

**Methods**

**Samples preparation**: A No. 1 (0.13 - 0.16 mm thick) cover slip was taken and carefully washed multiple times with acetone and isopropanol in a clean room environment. Layers of titanium 1 nm, gold 88 nm and chromium 1 nm thick were evaporated sequentially onto the cover slip using an Electron Beam Evaporator (EBE, Lesker PVD). A thin layer of Titanium is deposited to ensure adhesion of Gold to the glass substrate. A Focused Ion Beam (FIB, FEI NanoLab 600 dual beam system) milling at 80 pA current and 30 kV voltage was used to drill elliptical holes with diameters of 130 nm and 80 nm on the long and short axis lengths respectively. The upper layer of chromium was removed using a Cr etch after removing the cover slips from the EBE and before performing SNOM measurements on them. Chromium layer is deposited and then etched to give straight edges to the nanoholes after the milling process.

**Samples characterization**: The far-field absorption spectrum of the Au thin film (see Fig.6a) has been measured by a UV/VIS/NIR spectrometer (Perkin-Elmer Lambda 2) and shows a broad band from red to infrared region. AFM (Atomic Force Microscopy) images of the sample surface were acquired with a NT-MDT NTEGRA Spectra microscope working in semi-contact mode. Figure 6b shows the AFM image with a scan area of 30μm×30μm. The sample morphology is characterized by a smooth surface with nanoholes placed at distance of 14μm from each other. This distance is large enough to ensure that nanoholes are optically isolated. Furthermore, nanoholes are characterized by an ellipsoidal shape (see AFM image with a scan area of 5μm×5μm in Fig.6c) with a major axis of about 130nm and a minor axis of about 80nm.

Figure 1d shows SNOM (Scanning Near-field Optical Microscopy) setup used for near-field optical characterization of a single nanohole. SNOM measurements were performed in transmission mode with hollow-pyramid cantilever[49] working in contact mode. As exciting source, we used an unpolarized He-Ne laser coupled to a single mode optical fiber (THORLABS). The wavelength of laser source ($\lambda_{exc} = 632$ nm) falls inside the absorption band of the metal film and therefore it is able to excite the SPPs. In order to define a polarization direction for the incident radiation, a linear polarizer (#1) is placed in the optical path between the end of the optical fiber and one aspherical lens. The polarized beam was focused onto the sample surface and moved on the plane of the sample by a micrometric *xyz* positioning system. The polarizer is mounted on a rotating support in order to control the polarization direction of the incident radiation and to selectively excite localized plasmon resonances (LPRs) related to minor or major or either axis of the ellipsoidal nanohole. The radiation pattern on the sample surface is collected in near-field by a SNOM hollow cantilever and by a 100X objective (Mitutoyo, NA = 0.70). The transmitted light passes through a second polarizer (#2) and, finally, it is detected by a cooled CCD Camera (Andor IDus). Also, the second polarizer is mounted on a rotating support in order to analyze the polarization of the collected radiation. In this way, it is possible to decide whether or not to exclude the incident electromagnetic field contribution and to observe the de-polarization process of light that goes through the elongated nanohole.

**Modelling**: The finite elements method (FEM) is a differential equation method that computes the scattered time-harmonic electromagnetic field by numerically solving the vector Helmholtz equation subject to boundary conditions at the particle surface. Due to the discretization process (mesh construction) of the domains in smaller parts, this technique is suitable to describe systems with complex geometry. The simulated structure is constituted of an Au thin film with a thickness of 88 nm with an elliptical nanohole placed at its center with major axis $a = 130$ nm and minor axis $b = 80$ nm. The entire structure is placed on a glass substrate and it is embedded in a 3D finite computational domain, discretized into many small-volume tetrahedral cells. In the far-field zone, at the outer boundary of the finite computational domain, approximate absorbing boundary conditions are imposed. In this way it is possible to suppress wave reflections back into the domain allowing the propagation of the numerical outgoing waves as if the domain were infinite. In the near-field zone, for the 3D tetrahedral elements, we choose an extremely fine mesh size (up to 0.1 nm) to better describe the rapid changes in the electromagnetic field. By solving 3D Helmholtz equations together with the boundary conditions in each element of the mesh, we can reconstruct the optical behavior of the system and the near-field enhancement distribution around the nanostructures.

# References

1. B. Hecht, B. Sick, U. P. Wild. Scanning near-field optical microscopy with aperture probes: fundamentals and applications. *The Journal of Chemical Physics* **112**, 7761 (2000).
2. D. J. Shin, A. Chavez-Pirson. Diffraction by a subwavelength-sized aperture in a metal plane. *J. Opt. Soc. Am. A* **18**, 1477 (2001).
3. A. Ambrosio, E. Cefalì, S. Spadaro, S. Patanè, M. Allegrini, D. Albert, E. Oesterschulze. Noncontact tuning fork position sensing for hollow-pyramid near-field cantilevered probes. *APL* **89**, 163108 (2006).
4. C. Triolo, A. Cacciola, R. Saija, S. Trusso, M. C. Spadaro, F. Neri, P. M. Ossi, S. Patanè. Near-field optical detection of Plasmon Resonance from Gold Nanoparticles: Theoretical and Experimental Evidence. *Plasmonics* **10**, 63-70 (2015).
5. P. G. Gucciardi, S. Trusso, C. Vasi, S. Patanè, M. Allegrini. Nano-Raman imaging of Cu-TCNQ clusters in TCNQ thin films by scanning near-field optical microscopy. *Phys. Chem. Chem. Phys.* **4**, 2747-2753 (2002).
6. C. Triolo, S. Patanè, M. Mazzeo, S. Gambino, G. Gigli, M. Allegrini. Pure optical nano-writing on light-switchable spiropyrans/merocyanine thin film. *Optics Express* **22**, 283-288 (2014).
7. A. Ambrosio, M. Alderighi, M. Labardi, L. Pardi, F. Fuso, M. Allegrini, S. Nannizzi, A. Pucci, G. Ruggeri. Near-field optical microscopy of polymer-based films with dispersed terthiophene chromophores for polarizer applications. *Nanotechnology* **15**, S270 (2004).
8. T. Matsui, A. Agrawal, A. Nahata, Z. V. Vardeny. Transmission resonances through aperiodic arrays of subwavelength apertures. *Nature* **446**, 517 (2007).
9. C. Genet, T. W. Ebbesen. Light in tiny holes. *Nature Review* **445**, 39 (2007).
10. A. Ivinskaya, M. I. Petrov, A. A. Bogdanov, I. Shishkin, P. Ginzburg, A. S. Shalin. Plasmon-assisted optical trapping and anti-trapping. *Light: Science & Applications* **6**, (2017).
11. C. Manzo, T. S. van Zanten, M. F. Garcia-Parajo. Nanoscale Fluorescence Correlation Spectroscopy on Intact Living Cell Membranes with NSOM Probes. *Biophysical Journal* **100**, L08 (2011).
12. O. M. Maragò, P. H. Jones, P. G. Gucciardi, G. Volpe, A. C. Ferrari. Optical trapping and manipulation of nanostructures. *Nature Nanotechnology* **8**, 807 (2013).
13. J.-M. Yi, A. Cuche, F. de Leon-Pérez, A. Degiron, E. Laux, E. Devaux, C. Genet, J. Alegret, L. Martìn-Moreno, T. W. Ebbesen. Diffraction Regimes of Single Holes. *PRL* **109**, 023901 (2012).
14. Bethe, H. A. Theory of diffraction by small holes. *The Physical Review* **66**, 143 (1944).
15. A. Degiron, H.J. Lezec, N. Yamamoto, T.W. Ebbesen. Optical Transmission properties of a single subwavelength aperture in a real metal. *Optics Communications* **239**, 61-66 (2004).
16. K. Y. Bliokh, F. Nori. Transverse spin of a surface polariton. *Phys. Rev. A* **85**, 061801(R) (2012).
17. Y. Gorodetski, A. Niv, V. Kleiner, E. Hasman. Observation of the Spin-Based Plasmonic Effect in Nanoscale Structures. *PRL* **101**, 043903 (2008).
18. K. Y. Bliokh, F. J. Rodríguez-Fortuño, F. Nori, A. V. Zayats. Spin–orbit interactions of light. *Nature Photonics* **9**, 796 (2015).
19. X. Yin, Z. Ye, J. Rho, Y. Wang, X. Zhang. Photonic Spin Hall Effect at Metasurfaces. *Science* **339**, 1405 (2013)
20. N. Shitrit, I. Bretner, Y. Gorodetski, V. Kleiner, E. Hasman. Optical Spin Hall Effects in Plasmonic Chains. *Nano Lett.* **11**, 2038-2042 (2011).
21. X. Ling, X. Zhou, K. Huang, Y. Liu, C.-W. Qiu, H. Luo, and S. Wen. Recent advances in the spin Hall effect of light. *Rep. Prog. Phys.* **80**, 066401 (2017).
22. F. Bouchard, I. De Leon, S. A. Schulz, J. Upham, E. Karimi, R. W. Boyd. Optical spin-to-orbital angular momentum conversion in ultra-thin metasurfaces with arbitrary topological charges. *Appl. Phys. Lett.* **105**, 101905 (2014).
23. R. C. Devlin, A. Ambrosio, N. A. Rubin, J. P. B. Mueller, F. Capasso. Arbitrary spin-to-orbital angular momentum conversion of light. *Science* **358**, 896-901 (2017).
24. J. Olmos-Trigo, J. J. Saenz. Spin control of macroscopic objects. *Nature Photonics* **12**, 444-450 (2018).
25. S.-W. Cho, J. Park, S.-Y. Lee, H. Kim, B. Lee. Coupling of spin and angular momentum of light in plasmonic vortex. *Optics Express* **9**, 10083 (2012).
26. C.-F. Chen, C.-T. Ku, Y.-H. Tai, P.-K. Wei, H.-N. Lin, C.-B. Huang. Creating Optical Near-Field Orbital Angular Momentum in a Gold Metasurface. *Nano Lett.* **15**, 2746-2750 (2015).
27. C. Triolo, A. Cacciola, S. Patanè ,R. Saija, S. Savasta, F. Nori. Spin-Momentum Locking in the Near Field of Metal Nanoparticles. *ACS Photonics* **4**, 2242-2249 (2017).
28. K. Y. Bliokh, A. Y. Bekshaev, F. Nori. Extraordinary momentum and spin in evanescent waves. *Nature Communications* **5**, 3300 (2014).
29. C. Leyder, M. Romanelli, J. Ph. Karr, E. Giacobino, T. C. H. Liew, M. M. Glazov, A. V. Kavokin, G. Malpuech, A. Bramati. Observation of the optical spin Hall effect. *Nature Physics* **3**, 628 (2007).
30. D. O'Connor, P. Ginzburg, F. J. Rodríguez-Fortuño, G. A. Wurtz, and A. V. Zayats. Spin–orbit coupling in surface plasmon scattering by nanostructures. *Nat. Commun.* **5**, 5327 (2014).

31. M. Antognozzi, C.R. Bermingham, R.L. Harniman, S. Simpson, J. Senior, R. Hayward, H. Hoerber, M.R. Dennis, A.Y. Bekshaev, K.Y. Bliokh, F. Nori. Direct measurements of the extraordinary optical momentum and transverse spin-dependent force using a nano-cantilever. *Nature Physics* **12**, 731 (2016).
32. K.Y. Bliokh, A.Y. Bekshaev, F. Nori. Optical Momentum, Spin, and Angular Momentum in Dispersive Media. *Phys. Rev. Lett.* **119**, 073901 (2017).
33. M.F. Picardi, K.Y. Bliokh, F.J. Rodriguez-Fortuno, F. Alpeggiani, F. Nori. Angular momenta, helicity, and other properties of dielectric fiber and metallic-wire modes. *Optica* **5**, 1016 (2018).
34. Y.-H. Wang, R.-C. Jin, J.-Q. Li, F. Zhong, H. Liu, I. Kim, Y. Jo, J. Rho, Z.-G. Dong. Photonic spin Hall effect by the spin-orbit interaction in a metasurface with elliptical nano-structures. *Appl. Phys. Lett.* **110**, 101908 (2017).
35. A. Bekshaev, K.Y. Bliokh, M. Soskin. Internal flows and energy circulation in light beams. *J. Opt.* **13**, 053001 (2011).
36. L. T. Vuong, A. J. L. Adam, J. M. Brok, P. C. M. Planken, H. P. Urbach. Electromagnetic Spin-Orbit Interactions via Scattering of Subwavelength Apertures. *PRL* **104**, 083903 (2010).
37. J.-J. Greffet, R. Carminati. Image Formation in Near-Field Optics. *Progress in Surface Science* **56**, 133-237 (1997).
38. A. Martinez, I.I. Smalyukh. Light-driven dynamic Archimedes spirals and periodic oscillatory patterns of topological solitons in anisotropic soft matter. *Opt. Express* **23**, 4591 (2015).
39. C.-T. Ku, H.-N. Lin, C.-B. Huang. Direct observation of surface plasmon vortex and subwavelength focusing with arbitrarily-tailored intensity patterns. *APL* **106**, 053112 (2015).
40. T. Rindzevicius, Y. Alaverdyan, B. Sepulveda, T. Pakizeh, M. Kall. Nanohole Plasmons in Optically Thin Gold Films. *J. Phys. Chem. C* **111**, 1207-1212 (2007).
41. M. R. Dennis, K. O'Holleran, M. J. Padgett. Singular Optics: Optical Vortices and Polarization Singularities. *Progress in Optics* **53** (2009).
42. K. Y. Bliokh, F. Nori. Transverse and longitudinal angular momenta of light. *Physics Reports* **592**, 1-38 (2015).
43. M.S. Soskin, M.V. Vasnetsov. Singular optics. *Prog. Opt.* **42**, 219-276 (2001).
44. S. Savasta, R. Saija, A. Ridolfo, O. Di Stefano, P. Denti, F. Borghese. Nanopolaritons: Vacuum Rabi Splitting with a Single Quantum Dot in the Center of a Dimer Nanoantenna. ACS Nano 4, 6369–6376 (2010).
45. A. Cacciola, O. Di Stefano, R. Stassi, R. Saija, S. Savasta. Ultrastrong Coupling of Plasmons and Excitons in a Nanoshell. *ACS Nano* **8**, 11483–11492 (2014).
46. A. Cacciola, C. Triolo, O. Di Stefano, A. Genco, M. Mazzeo, R. Saija, S. Patanè, S. Savasta. Subdiffraction Light Concentration by J-Aggregate Nanostructures. *ACS Photonics* **2**, 971-979 (2015).
47. B. H. Woo, I. C. Seo, E. Lee, S. Y. Kim, T. Y. Kim, S. C. Lim, H. Y. Jeong, C. K. Hwangbo, Y. C. Jun. Dispersion Control of Excitonic Thin Films for Tailored Superabsorption in the Visible Region. *ACS Photonics* **4**, 1138-1145 (2017).
48. A. Ridolfo, R. Saija, S. Savasta, P.H. Jones, M.A. Iatì, O.M. Maragò. Fano-Doppler Laser Cooling of Hybrid Nanostructures, *ACS Nano* **5**, 7354–7361 (2011).
49. P. N. Minh, T. Ono, S. Tanaka, M. Esashi. Spatial distribution and polarization dependence of the optical near-field in a silicon microfabricated probe. *Journal of Microscopy* **202**, 28-33 (2000).

**Author contributions**

S.P., S.S. and C.T. have developed the concepts. N-R. A. and S. T. have prepared the samples. C.T. and S.P. have performed all measurements. C.T. and R.S. have performed the simulations. All authors contributed to writing the manuscript.

**Competing interests**: The authors declare no competing interests.

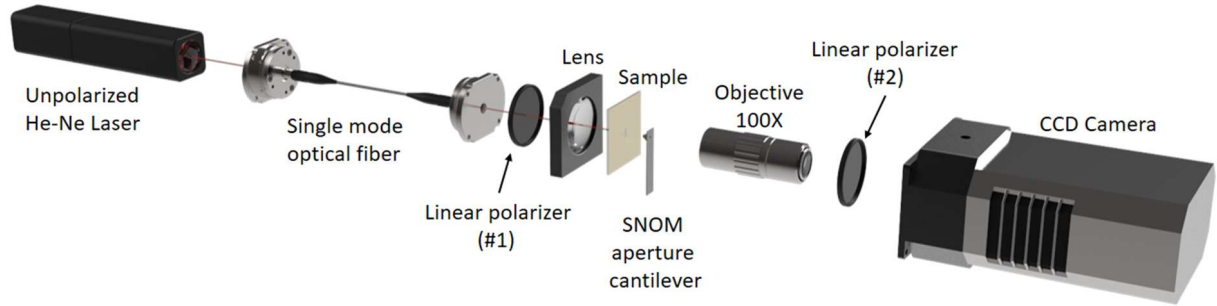

**Figure 1.** SNOM (Scanning Near-field Optical Microscopy) setup used for near-field characterization of single nanohole. SNOM measurements are performed in transmission mode with a NT-MDT NTEGRA Spectra microscope using a SNOM aperture cantilever in contact mode. As exciting source, we used an unpolarized He-Ne laser ($\lambda_{exc} = 632nm$) coupled to a single mode optical fiber. A linear polarizer (#1) is placed at the end of the optical fiber and the polarized beam is focused onto the sample surface. Resulting radiation is collected in near-field by a SNOM aperture cantilever and by a $100X$ objective (Mitutoyo, $NA = 0.70$). Transmitted light from the sample passes through a second polarizer (#2) and, finally, detected by a cooled CCD Camera (Andor IDus).

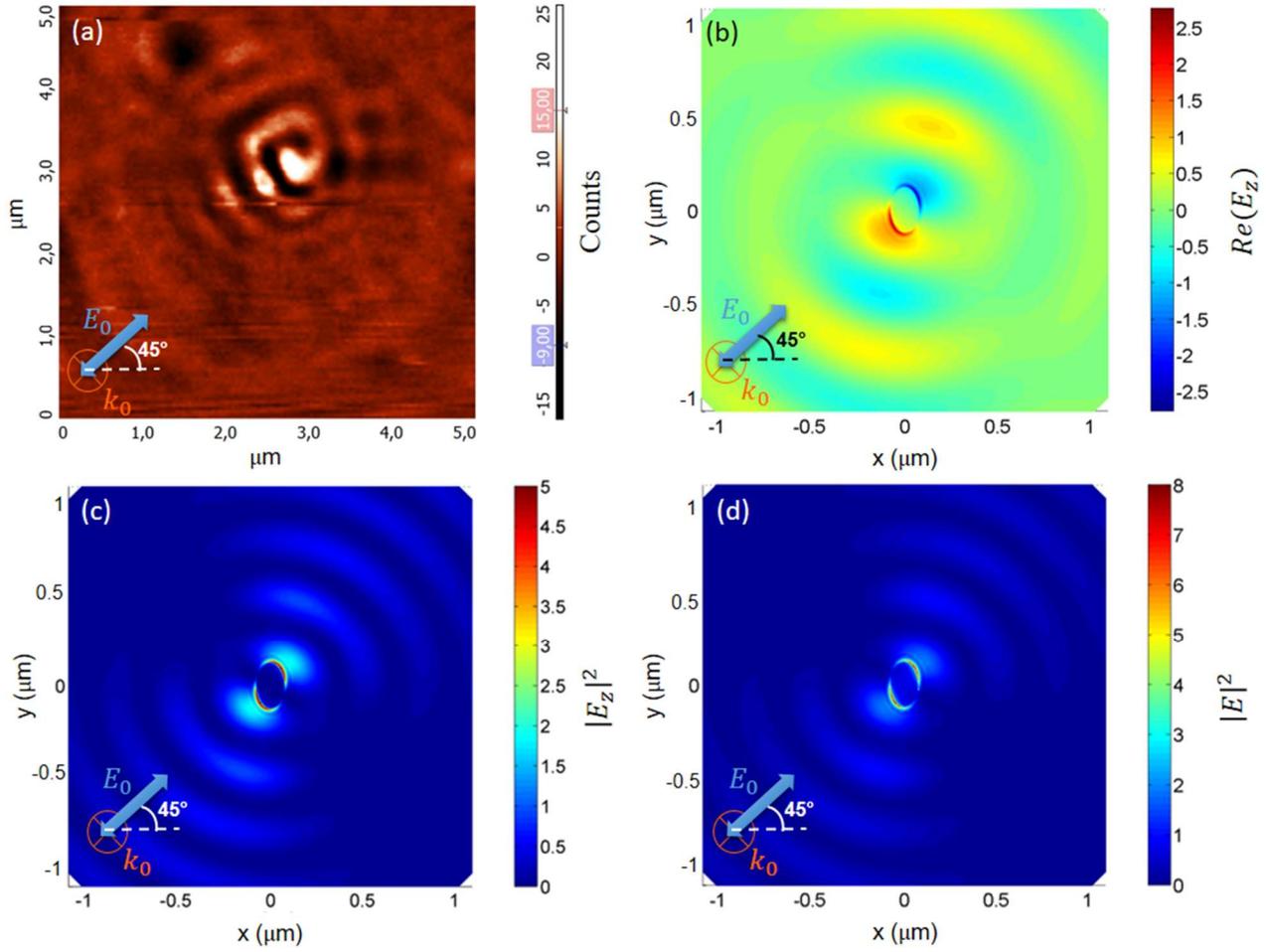

**Figure 2.** (a) SNOM image (scan area 5μm×5μm) performed using an incident beam at $\lambda_{exc} = 632$ nm with normal incidence and linear polarization at 45° on an elliptical nanohole in a thin gold film. The second polarizer is orthogonal respect to the first one (#1⊥#2). Numerical results: (b) Near-field intensity distribution of $Re(E_z)$ around the elliptical nanohole illuminated by a plane wave at λ=632 nm, at 2 nm from the metal surface. Simulations are performed by finite element method (FEM) simulation. Size of nanohole: major axis a=130 nm, minor axis b=80 nm. (c) Near-field intensity distribution of the z-component of the electric field $|E_z|^2$ around the elliptical nanohole illuminated by a plane wave at $\lambda_{exc} = 632$ nm, at 2 nm from the metal surface. (d) Near-field intensity distribution of the total electric field $|E|^2$ around the elliptical nanohole illuminated by a plane wave at $\lambda_{exc} = 632$ nm, at 2 nm from the metal surface.

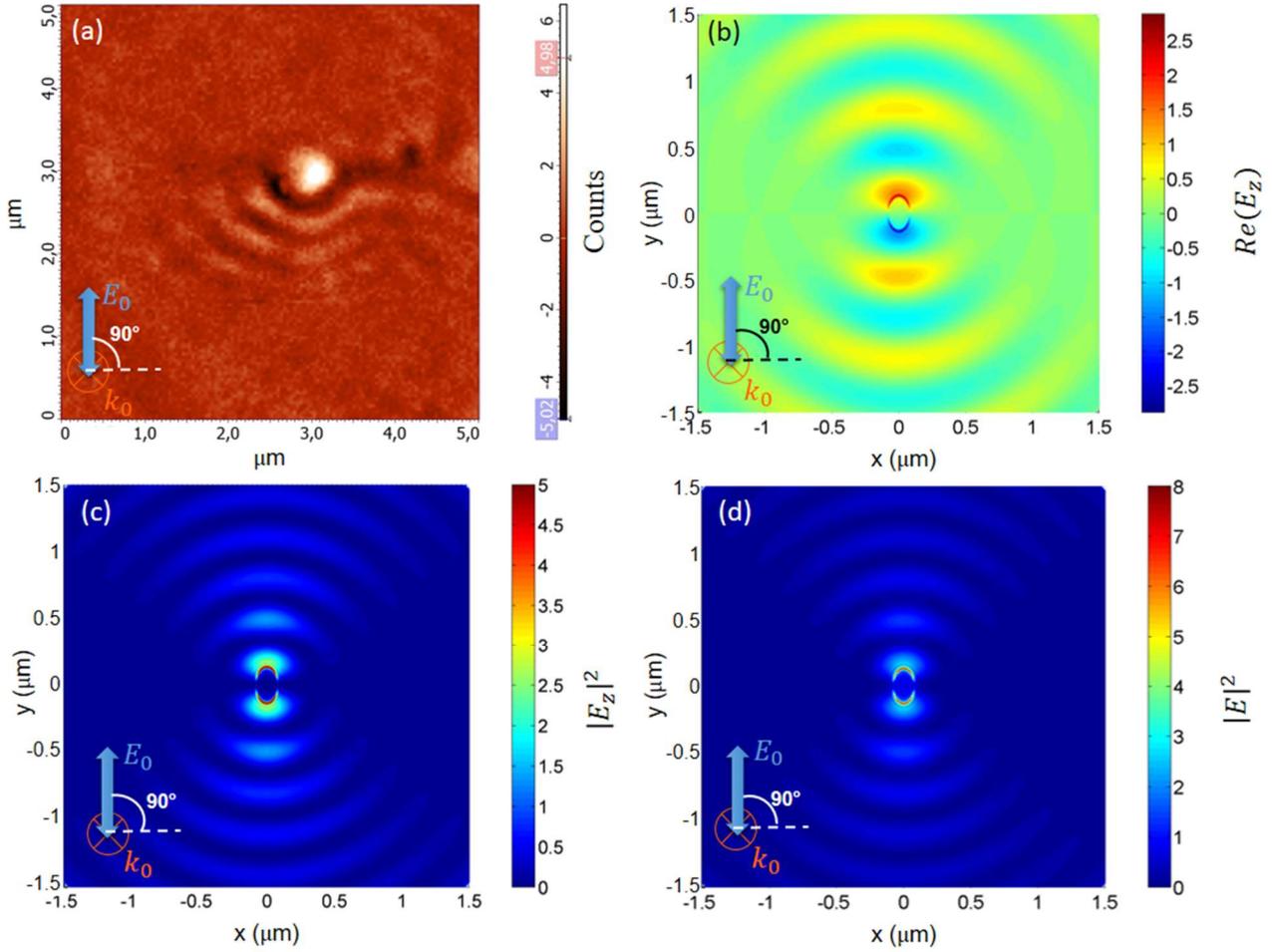

**Figure 3.** (a) SNOM image (scan area 5μm×5μm) performed using an incident beam at $\lambda_{exc} = 632$ nm with normal incidence and linear polarization at 90° on an elliptical nanohole in a thin gold film. The second polarizer is orthogonal respect to the first one (#1⊥#2). Numerical results: (b) Near-field intensity distribution of $Re(E_z)$ around the elliptical nanohole illuminated by a plane wave at λ=632 nm, at 2 nm from the metal surface. Simulations are performed by finite element method (FEM) simulation. Size of nanohole: major axis a=130 nm, minor axis b=80 nm. (c) Near-field intensity distribution of the z-component of the electric field $|E_z|^2$ around the elliptical nanohole illuminated by a plane wave at $\lambda_{exc} = 632$ nm, at 2 nm from the metal surface. (d) Near-field intensity distribution of the total electric field $|E|^2$ around the elliptical nanohole illuminated by a plane wave at $\lambda_{exc} = 632$ nm, at 2 nm from the metal surface.

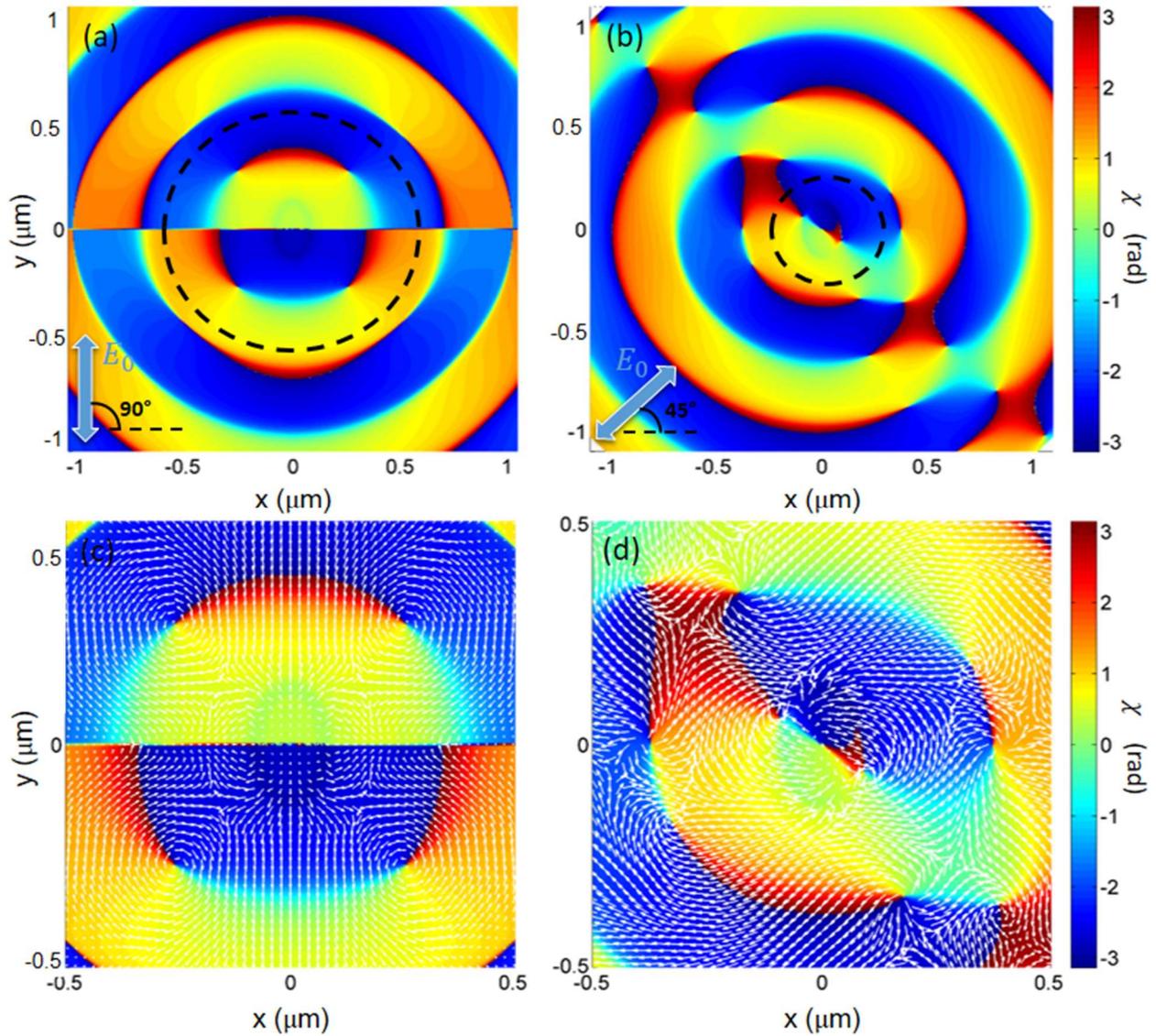

**Figure 4.** Numerical results: (a) Spatial distribution (2μm × 2μm) of the phase variation of the z-component of the total field calculated by FEM simulations at 2 nm from the Au surface for an incident polarization direction coincident with the major axis of the elliptical nanohole at normal incidence. The dashed black line encloses the region with the balance of total topological charge. (b) Spatial distribution (2μm × 2μm) of the phase variation of the z-component of the total field calculated by FEM simulations at 2 nm from the Au surface for an incident polarization direction rotated of 45° respect to the major axis of the elliptical nanohole at normal incidence. The dashed black line encloses the region with the unbalance of total topological charge. (c) Spatial distribution (1μm × 1μm) of the phase variation of the z-component of the total field calculated by FEM simulations at 2 nm from the Au surface for an incident polarization direction shifted of 45° respect to the major axis of the elliptical nanohole at normal incidence. (c) Spatial distribution (1μm × 1μm) of the phase variation of the z-component of the total field calculated by FEM simulations at 2 nm from the Au surface for an incident polarization direction coincident with the major axis of the elliptical nanohole at normal incidence. In figures (c) and (d), white arrows are the optical momentum vectors ($p_0$) that evidence the formation of optical vortices around the phase singularities.

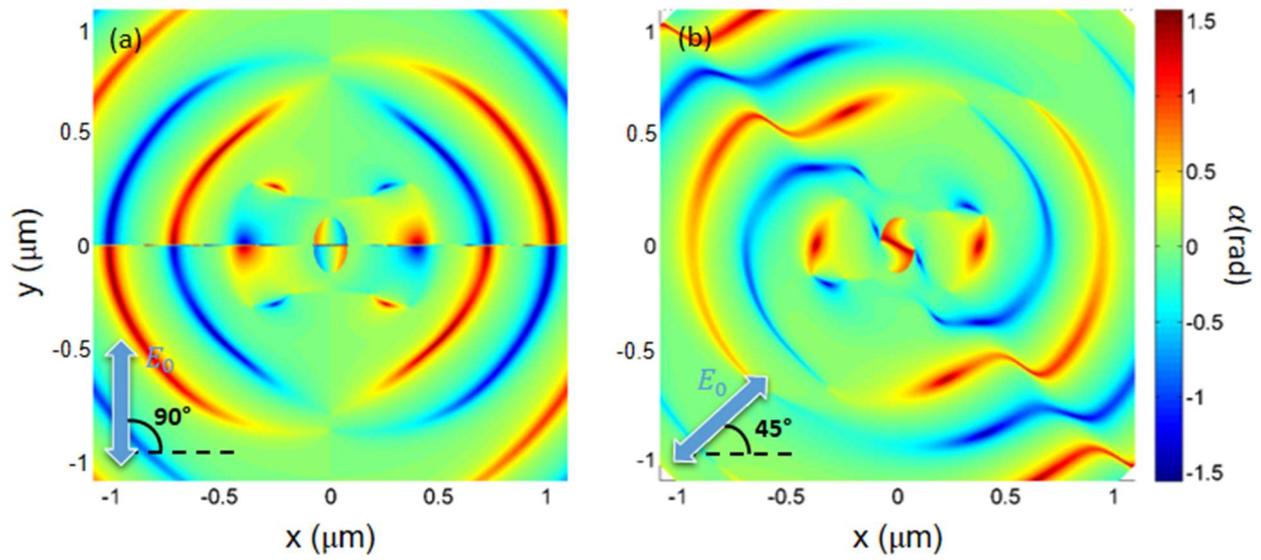

**Figure 5.** Numerical results: (a) Spatial distribution (2μm × 2μm) of the angle that spin vector $s$ forms with the $xy$-plane calculated by FEM simulations at 2 nm from the Au surface for an incident polarization direction coincident with the major axis of the elliptical nanohole at normal incidence. (b) Spatial distribution (2μm × 2μm) of the phase variation of the $z$-component of the total field calculated by FEM simulations at 2 nm from the Au surface for an incident polarization

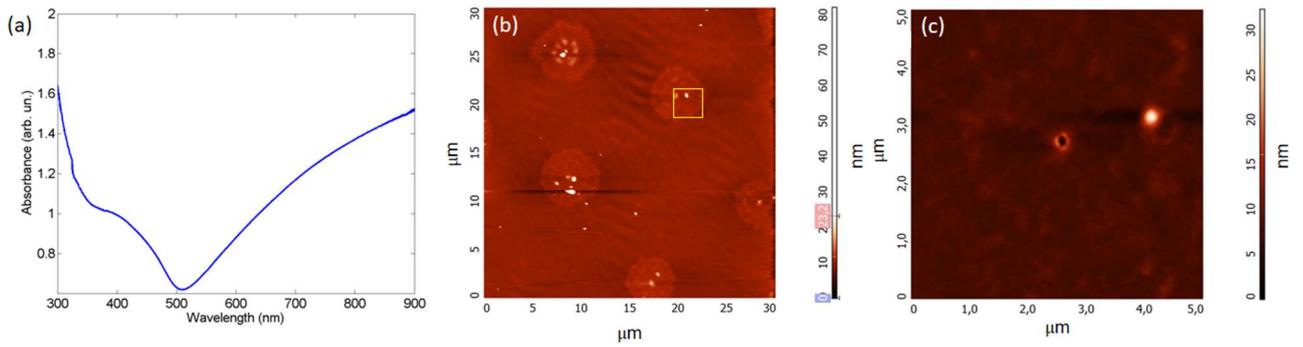

**Figure 6.** (a) Far-field absorption spectrum of the Au thin film that shows a broad absorption band from red to infrared region. (b) AFM image with a scan area of 30μm×30μm. Sample morphology is characterized by a smooth surface with nanoholes placed at distance of 14μm from each other. (c) Detail of the AFM image in (b) (highlighted with a yellow square) that shows an AFM image with a scan area of 5μm×5μm. All nanoholes are characterized by an ellipsoidal shape with a major axis of about 136nm and a minor axis of about 80nm.